\documentclass[aps,prl,twocolumn,superscriptaddress,showpacs]{revtex4}

\usepackage{graphicx}

\begin{document}

\title{Two-subband quantum Hall effect in parabolic quantum wells}

\author{C. Ellenberger}
\author{B. Simovi\v c}
\author{R. Leturcq}
\author{T. Ihn}
\author{S. E. Ulloa*}
\author{K. Ensslin}
\affiliation{Solid State Physics Laboratory, ETH Zurich, 8093 Zurich, Switzerland}
\author{D. C. Driscoll}
\author{A. C. Gossard}
\affiliation{Materials Department, University of California, Santa Barbara, CA 93106, USA}

\date{\today}

\begin{abstract}
The low-temperature magnetoresistance of parabolic quantum wells displays pronounced minima between integer filling factors. Concomitantly  the Hall effect exhibits overshoots  and plateau-like features next to well-defined ordinary quantum Hall plateaus. These effects set in with the occupation of the second subband. We discuss our observations in the context of single-particle Landau fan charts of a two-subband system empirically extended by a density dependent subband separation and an enhanced spin-splitting $g^\star$.
\end{abstract}


\maketitle

At low temperatures a two-dimensional electron gas subject to a perpendicular magnetic field gives rise to the integer quantum Hall effect \cite{Prange90}.
The Hall effect takes on quantized values $\rho_{xy}=h/e^2\nu$ around magnetic fields $B=N_\mathrm{s}h/e\nu$, with the integer filling factor $\nu$ counting the number of occupied Landau levels and $N_\mathrm{s}$ being the carrier density. In these plateau  regions of $\rho_{xy}$ the magnetoresistance $\rho_{xx}$ is exponentially suppressed.
Each Landau level is described by a Landau level quantum number $n=0,1,2,\ldots$, a spin quantum number $s=\pm1/2$ and a subband or well index $i$ if several subbands in a quantum well are occupied or a double well system is investigated. 

Zhang {\it et al.} \cite{Zhang05} have reported
measurements on a quantum well system where the occupation of a second subband 
leads to additional maxima and minima in $\rho_{xx}$ {\em between} integer filling factors, accompanied by over- and undershoots of $\rho_{xy}$. Here we report similar results obtained on {\em parabolic} quantum wells in which the subband occupation can be controlled by front and back gate voltages.
We find that the observed phenomena can be qualitatively accounted for using an effective single-particle model where interaction effects beyond the mean field Hartree and exchange interactions are neglected. Previous experiments on double wells showed additional minima as well as hysteretic behavior (see Ref.~\cite{Girvin00} for a review) which was attributed to more subtle interaction effects \cite{Jungwirth01,Jungwirth01a}.

Our samples are 100~nm wide parabolic Al$_x$Ga$_{1-x}$As quantum wells with the Al-content varying from $x=0.4$ at the edges to $x=0$ in the center of the parabola. Hall bars with Ohmic contacts were fabricated. A Schottky front gate and a back gate \cite{Lindemann02} allows to tune the electron sheet density $N_\mathrm{s}$ and with it the number of occupied subbands. At a temperature of 100 mK and zero gate voltages the electron mobility is $16$~m$^2$/Vs and $N_\mathrm{s}=3.3\times 10^{15}$~m$^{-2}$. While we have investigated many similar samples in the past \cite{Ensslin91,Ensslin93,Salis97,Salis01,Lindemann02} only one (Figs. 8 and 9 of Ref. \onlinecite{Ensslin93}) showed features which might be precursors of the  pronounced structures presented here. The only difference of the present compared to previous samples is the slightly higher mobility (10-20\%). Data was taken by two different experimentalists over a period of two years on 3 different samples from the same wafer with Hall bars of different size. All results are comparable and highly reproducible. In the following we show coherent data from one sample.

\begin{figure}
\includegraphics[width=3.4in]{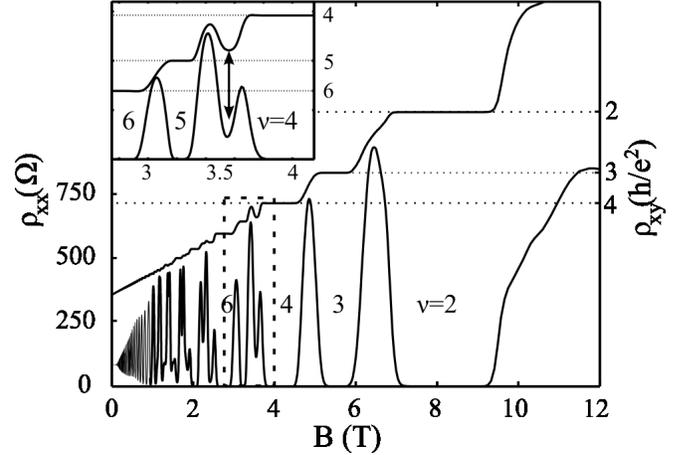}
\caption{Magnetoresistance $\rho_{xx}$ (left hand scale, lower curve) and Hall resistance $\rho_{xy}$ (right hand scale, upper curve) measured at T=100 mK. The horizontal lines indicate the position of the theoretically expected Hall plateaus. The inset in the upper left corner shows the same data but expanded in the regime of interest.\label{fig1}}
\end{figure}

Figure~\ref{fig1} shows typical experimental $\rho_{xx}$ and $\rho_{xy}$ traces when the second subband is populated. At magnetic fields $B<1\;\mathrm{T}$ Shubnikov-de Haas (SdH) oscillations are observed. The fast oscillation period is related to the lower subband density $N_\mathrm{s}^{(1)}=3.69\times 10^{15}$~m$^{-2}$ and the slow amplitude modulation to the upper subband density $N_\mathrm{s}^{(2)}=1.48\times 10^{14}$~m$^{-2}$. At $B>4\;\mathrm{T}$ standard SdH oscillations as well as quantum Hall (QH) plateaus occur corresponding to the total carrier density $N_\mathrm{s}^\mathrm{tot}=N_\mathrm{s}^{(1)}+N_\mathrm{s}^{(2)}=3.84\times 10^{15}$~m$^{-2}$.
The unusual behavior at $2\;\mathrm{T} < B < 4\;\mathrm{T}$ is enlarged in the inset of Fig.~\ref{fig1}. An additional minimum occurs in $\rho_{xx}$ between filling factors $\nu=4$ and 5, accompanied by an oscillatory feature in $\rho_{xy}$ (arrow in the inset). These unexpected experimental observations are the focus of this paper and will be presented in more detail in the following.

\begin{figure}
\includegraphics[width=3.4in]{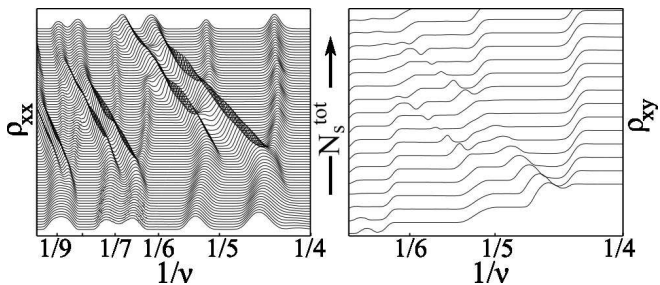}
\caption{Magnetoresistance $\rho_{xx}$ (left) and Hall resistance $\rho_{xy}$ (right)  at T=100 mK as a function of the inverse filling factor $1/\nu=eB / h N_\mathrm{s}^\mathrm{tot}$. Curves for different carrier densities $N_\mathrm{s}^\mathrm{tot}$ are vertically offset for clarity. The carrier density $N_\mathrm{s}^\mathrm{tot}$ was tuned between $3.4\times 10^{15}$~m$^{-2}$ and $4.85\times 10^{15}$~m$^{-2}$ (left) and between $3.77\times 10^{15}$~m$^{-2}$ and $4.85\times 10^{15}$~m$^{-2}$ (right).\label{fig2}}
\end{figure}

Figure~\ref{fig2} shows the magnetoresistance (left) and Hall resistance (right) for a data set where the carrier density $N_\mathrm{s}^\mathrm{tot}$ as determined from the classical low-field Hall effect was tuned using the front gate. By plotting the data as a function of $1/\nu$ features related to the same filling factor line up vertically. The magnetoresistance maximum between $\nu=4$ and $5$ splits as a function of increasing $N_\mathrm{s}$, crosses the $\nu=5$ and 6 valleys and joins the maximum  between $\nu=6$ and $7$. Similarly the maximum between $\nu=5$ and $6$ moves into the maximum between $\nu=7$ and $8$. Weaker features with qualitatively similar behavior are observed for the lower two sets of inverse filling factors. As one follows, e.g., the $\nu=5$ minimum, for low carrier densities (curves on the bottom of the figure) a single minimum is observed. As the carrier density is increased a double minimum feature (i.e., an additional maximum) occurs which then returns to a single minimum feature once the crossing regime is passed.

The Hall effect $\rho_{xy}$ shown on the right of Fig.~\ref{fig2} shows an oscillatory feature between two quantum Hall plateaus if $\rho_{xx}$ shows an additional maximum. In general the derivative $d\rho_{xy}/dB$ always behaves very similar to $\rho_{xx}$. Therefore we will restrict our discussion in the following to the behavior of $\rho_{xx}$.

\begin{figure}
\includegraphics[width=3in]{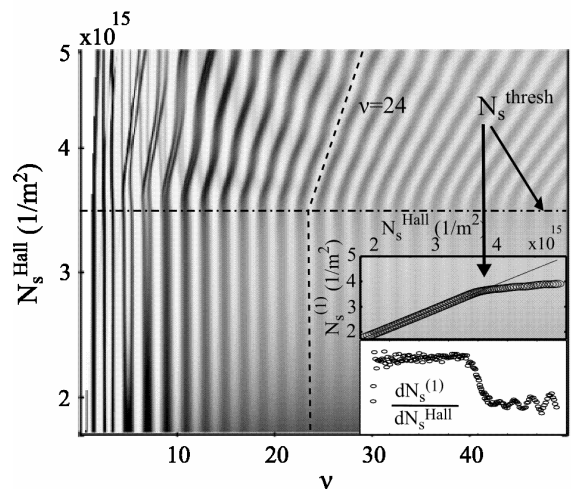}
\caption{Greyscale image of the magnetoresistance $\rho_{xx}$ versus filling factor calculated from the density extracted from the low-field Hall resistance and the applied magnetic field. The horizontal dash-dotted line marks the population onset of the second subband. The vertical dashed line follows the behavior of filling factor $\nu=24$. This dashed line is converted to a density and plotted in the top inset versus the Hall density. The lower inset shows the derivative $dN_\mathrm{s}^{(1)}/dN_\mathrm{s}^\mathrm{Hall}$ of the data in the upper inset highlighting the threshold density and revealing additional oscillations in $N_\mathrm{s}^{(1)}$.
 \label{fig3}}
\end{figure}

In order to define the parameter space more precisely we present low-field $\rho_{xx}$ data in a grey scale plot in Fig.~\ref{fig3} as a function of filling factor.
The vertical axes corresponds to the electron density $N_\mathrm{s}^\mathrm{Hall}$ as determined from the classical Hall effect at magnetic fields below 0.5~T, tuned with a front gate voltage. For densities below $N_\mathrm{s}^\mathrm{Hall}<N_\mathrm{s}^\mathrm{thresh} \approx 3.5 \times 10^{15}$~m$^{-2}$ the $\rho_{xx}$-minima corresponding to bright lines in the grey scale plot run vertically. This corresponds to the situation where one subband is occupied and the density $N_\mathrm{s}^{(1)}$ obtained from SdH oscillations agrees with the Hall density $N_\mathrm{s}^\mathrm{Hall}$ within measurement accuracy. At the threshold density of $N_\mathrm{s}^\mathrm{thresh} \approx 3.5 \times 10^{15}$~m$^{-2}$ the second subband becomes populated and therefore the total density measured via the Hall effect increases more strongly than $N_\mathrm{s}^{(1)}$ (see upper inset). This manifests itself by a sudden bending of the bright lines indicated, e.g., by the dashed line following filling factor $\nu_1=24$ for the lower subband. This way the carrier densities of both subbands $N_\mathrm{s}^{(1)}$ and $N_\mathrm{s}^{(2)}=N_\mathrm{s}^\mathrm{tot}-N_\mathrm{s}^{(1)}=N_\mathrm{s}^\mathrm{Hall}-N_\mathrm{s}^{(1)}$ can be determined \cite{Ensslin93,Salis97}. The solid line in the upper inset of Fig.~\ref{fig3} shows the relation $N_\mathrm{s}^\mathrm{tot}=N_\mathrm{s}^\mathrm{Hall}$. The dashed line in the main figure related to $\nu_1=24$ is plotted by circles and basically follows the solid line until the second subband becomes populated for $N_\mathrm{s}^\mathrm{tot} \le N_\mathrm{s}^\mathrm{thresh}$. For low enough magnetic fields, where the cyclotron energy $\hbar \omega_c$ is much smaller than the Fermi energy $E_F$, magnetic field induced redistribution of charges between the two subbands is small \cite{Ensslin93}. Only weak oscillations can be detected in $dN_\mathrm{s}^{(1)}/dN_\mathrm{s}^\mathrm{Hall}$ (lower inset of Fig.~\ref{fig3}) which become more pronounced for higher magnetic fields. We also note that $N_\mathrm{s}^\mathrm{Hall}$ as a function of front gate voltage increases its slope at $N_\mathrm{s}^\mathrm{thresh}$ due to the increased density of states at the Fermi energy.

\begin{figure}
\includegraphics[width=2.8 in]{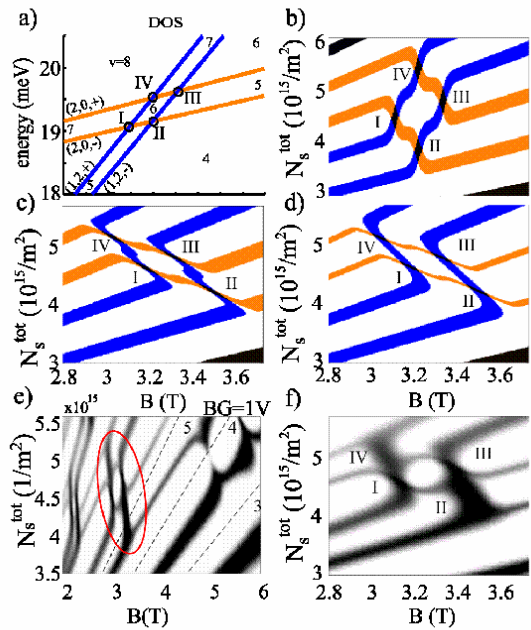}
\caption{(Color online) (a) Two spin-split Landau levels belonging to different subbands cross. (b) The same Landau levels as in (a) plotted in the $N_\mathrm{s}^\mathrm{tot}$--$B$ plane. (c) Crossing of the same Landau levels assuming a density dependent subband spacing $\Delta E$. (d) Crossing of the Landau levels as in (c) but allowing for different Landau-level broadenings $\Gamma_1$ and $\Gamma_2$ in the two subbands. (e) Measured $\rho_{xx}$ in the $N_\mathrm{s}^\mathrm{tot}$--$B$ plane. The encircled region in magnified in (f). Corresponding crossing points I--IV are labeled consistently in (a)--(f).
\label{fig5}}
\end{figure}

In order to check the influence of the confining potential we have measured
the magnetoresistance $\rho_{xx}$
for different values of the back gate voltage while tuning the carrier density via the front gate. A particular example is shown in Fig.~\ref{fig5}(e).
All these measurements show the same dominant ring-like features at the same density values. We conclude that it is the density which is responsible for the onset of the additional features in the magnetoresistance.

Data has also been taken for magnetic fields tilted with respect to the sample normal. If the diamagnetic shift resulting from the in-plane field component is considered the data look again very similar to Fig.~\ref{fig5}(e). For very high in-plane magnetic fields $B_\mathrm{in-plane} > 4T$ the threshold density $N_\mathrm{s}^\mathrm{thresh}$ where the second subband becomes populated can no longer be reached without significant parallel conduction in the doping layer. From the available data we have no indication that spin effects, which are expected to become more dominant for large in-plane fields, become more pronounced.

Temperature dependent experiments show that the additional minima disappear at lower temperatures than the nearby ordinary quantum Hall minima similar to the data reported in Ref.~\onlinecite{Zhang05}. This indicates that the related gaps in the energy spectrum are smaller than the Landau gap $\hbar \omega_c$. This gap is related to the spin-gap of spin-resolved Landau levels (see below). Data taken for different orientations of the Hall geometry with respect to the crystal axes are very similar excluding anisotropies of the underlying crystal lattice to be the origin for the observed phenomena. Within experimental resolution no hysteretic behavior in magnetic field and gate voltage was observed.

The data presented in Fig.~\ref{fig5}(e) and the detail in Fig.~\ref{fig5}(f) is generally similar to the results of Ref.~\onlinecite{Zhang05}. We interpret a dip in the magnetoresistance as being related to a gap in the energy spectrum of the system.
The general strength of the observed features increases with magnetic field and density (mobility). In a single-particle picture the lowest density at which the features presented in Fig.~\ref{fig5}(e) occur corresponds to the crossing points of the lowest Landau level of the upper subband with respective Landau levels in the lower subband. The total density can not be made large enough to observe the crossing points of the second Landau level in the upper subband with Landau levels in the lower subband. The lowest magnetic field, at which the features in Fig.~\ref{fig5}(e) can be detected generally coincides with the beginning observation of spin splitting in the lowest subband. We like to note, however, that weak precursors of such features can also be observed at relatively low magnetic fields where spin splitting can not yet be resolved (c.f. Fig.~\ref{fig3}). In this case there are no additional zeros in $\rho_{xx}$.

In the framework of filling factors related to the total carrier density $N_\mathrm{s}^\mathrm{tot}=N_\mathrm{s}^{(1)}+N_\mathrm{s}^{(2)}$ indicated by the dashed lines in Fig.~\ref{fig5}(e) there is one dark feature (i.e. maximum) crossing a magnetoresistance minimum for odd filling factors. In contrast there are two dark features (maxima) for even filling factors delineating the ring-like structures described in Ref. \onlinecite{Zhang05}. 

A qualitative picture emerges by considering Landau level crossings as schematically indicated in Fig.~\ref{fig5}(a). We label individual levels by $(i,n,s)$, i.e., the subband quantum number $i=1,2$, the Landau-level quantum number $n=0,1,2,\ldots$, and the spin quantum number $s=\pm 1/2$. The four involved Landau-levels are ($1,2,+$) and ($1,2,-$) with energies $E_{1,2,\pm} = 5\hbar\omega_\mathrm{c}/2\pm g^\star\mu_\mathrm{B}B$ and $(2,0,+)$ and $(2,0,-)$ with energy $E_\pm =\Delta E+ \hbar\omega_\mathrm{c}/2\pm g^\star\mu_\mathrm{B}B$. Here we have introduced the subband-separation in energy, $\Delta E$, $\omega_\mathrm{c}=eB/m^\star$ is the cyclotron frequency, $\mu_\mathrm{B}$ the Bohr-magneton and $g^\star$ the effective Land\'{e}-factor. The crossing region of these four energy levels corresponds to the experimentally observed region encircled in Fig.~\ref{fig5}(e).

In order to compare to the experiment, a crucial step is to translate the Landau-fan in the energy--magnetic field plane [Fig.~\ref{fig5}(a)] to the experimentally relevant plane spanned by $N_\mathrm{s}^\mathrm{tot}$ and magnetic field as shown in Fig.~\ref{fig5}(b). This is a highly non-linear mapping. We achieve this mapping by assuming Gaussian broadened Landau levels of width $\Gamma$. In transport experiments a peak in $\rho_{xx}$ is expected when the Fermi energy lies in extended states near the density of state peak of a Landau level. In Fig.~\ref{fig5}(b) we therefore color regions corresponding to a band of 30~$\mu$eV around density of states maxima. This (arbitrary) choice of threshold was checked not to modify the results on a qualitative level.

While in the Landau-fan [Fig.~\ref{fig5}(a)] the spacing between spin-split levels is typically different from the spacing of different Landau levels, in the $N_\mathrm{s}^\mathrm{tot}$ representation all density of states peaks (and therefore maxima in $\rho_{xx}$) are equally spaced, if the Landau level separation is much larger than $\Gamma$, because each Landau level can maximally be occupied by $eB/h$ electrons per area. If the Landau level separation in energy is comparable to their width, an uneven separation results.
Therefore, for narrow Landau levels and far away from degeneracy points, peaks of the density of states are found at densities $N_\mathrm{s}^\mathrm{tot}=eB/h(p+1/2)$ with integer numbers $p$ counting the number of completely filled Landau levels below the one considered [c.f. Fig.~\ref{fig5}(b)]. In contrast, at magnetic fields where Landau-levels of the two subbands cross (points I--IV), peaks in the density of states are found at $N_\mathrm{s}^\mathrm{tot}=q eB/h$
with integer numbers $q$, i.e., exactly in the middle between the lines formed by the non-degenerate states. This argument intuitively explains the behavior of the four Landau-levels presented in Fig.~\ref{fig5}(a) when translated into the $N_\mathrm{s}^\mathrm{tot}$--$B$ plane [Fig.~\ref{fig5}(b)]. Crossing points I--IV between the four levels occur at the same magnetic field values in Figs.~\ref{fig5}(a) and (b).
We assign the same crossing points to the measured data in Fig.~\ref{fig5}(f) which is a zoom into the region encircled in Fig.~\ref{fig5}(e).

In the data presented in Fig.~\ref{fig5}(f) we find slightly uneven spacing between the four levels at a given magnetic field far from the degeneracy points (e.g. at 2.8~T) indicating that the Landau-level width is only slightly smaller than the Zeeman splitting.
From crossing points II and IV we determine the subband separation $\Delta E = 2\hbar\omega_\mathrm{c}=11.7$~meV for point II and $10.5$~meV for point IV. We conclude that the subband spacing in the experiment depends on the total density, a finding which is in agreement with self-consistent calculations of parabolic (and also hard-wall) quantum wells \cite{Ensslin93}. From crossing points I and III we estimate $g^\star = \Delta E/2\mu_\mathrm{B}(1/B_\mathrm{I}-1/B_\mathrm{III})=2.1$, a value which is enhanced over the bulk value presumably due to exchange effects \cite{Ando82}.

Stimulated by the above findings we extended the simple Landau-fan model presented above by including a linearly density-dependent subband separation $\Delta E(N_\mathrm{s}^\mathrm{tot})$ and the experimentally determined enhanced $g^\star=2.1$ in Fig.~\ref{fig5}(c). It is remarkable that the emerging ring-like feature around 3.2 T resembles closely our data in Fig.~\ref{fig5}(f) as well as the data in Fig. 2 of Ref.~\onlinecite{Zhang05} in a single-particle picture, empirically extended by the density-dependent subband separation and the enhanced $g^\star $.

Even better agreement with our data in Fig.~\ref{fig5}(f) is achieved in Fig.~\ref{fig5}(d) where we allow for a subband dependent Landau-level broadening $\Gamma_i$. The value of $\Gamma_1$ was estimated from low-field SdH oscillations where we find $\hbar/\tau_\mathrm{q}=240$~$\mu$eV.
The value of $\Gamma_2$ was chosen in order to reproduce the experimentally observed motion of the resistivity-peaks shown in Fig.~\ref{fig5}(f).
A more detailed calculation should replace the empirical inclusion of $\Delta E(N_\mathrm{s}^\mathrm{tot})$ by a self-consistent treatment and the empirically enhanced $g^\star$ by the calculated magnetic field dependent quantity for both subbands \cite{Ando82}.
The subband density and with it the band structure will then become magnetic field-dependent as observed in the data in the lower inset of Fig.~\ref{fig3}. Possible interaction effects as discussed in Refs.~\onlinecite{Jungwirth01} and \onlinecite{Wang03} beyond those included above may further modify the details of the density of states. The basic experimental observation, however, can be explained based on the Landau level structure obtained from our effective single-particle model.

In conclusion, we have observed novel features in the magnetoresistance and Hall resistance of a two-subband system in a parabolic quantum well. The additional minima in $\rho_{xx}$ are remarkably pronounced, non-hysteretic and reproducible. We demonstrate that the dominant features of the experimental data can be explained by a single-particle model empirically extended by an enhanced $g^\star$ and a density-dependent subband separation. Evidence for interaction effects beyond this mean field Hartree and exchange approach cannot be reliably extracted from our data leaving this topic a challenge for future experiments.

\begin{acknowledgments}
The authors thank A. H. MacDonald and V. Falko for valuable discussions. Financial support from the Swiss Science Foundation (Schweizerischer Nationalfonds) and from the EU Human Potential Program financed via the Bundesministerium f{\"u}r Bildung und Wissenschaft is gratefully acknowledged.
\end{acknowledgments}

*: permanent address: Department of Physics and Astronomy, Ohio University, Athens, Ohio 45701-2979, USA 


\end{document}